\documentclass[aps,prb,showpacs,twocolumn,superscriptaddress]{revtex4}
\usepackage[english]{babel}
\usepackage{graphicx}
\usepackage{color}
\usepackage[utf8]{inputenc}
\usepackage{pstricks,pst-grad,color}
\usepackage{graphicx,pifont,amssymb}
\usepackage{amsmath,amssymb}

\begin{document}


\title{The surface density of states of layered {\it f}-electron materials}



\author{Robert Peters}
\email[]{peters@scphys.kyoto-u.ac.jp}
\affiliation{Department of Physics, Kyoto University, Kyoto 606-8502,
  Japan}

\author{Norio Kawakami}
\affiliation{Department of Physics, Kyoto University, Kyoto 606-8502,  Japan}

\date{\today}

\begin{abstract}
We theoretically analyze the surface density of states of heavy fermion materials such as CeCoIn$_5$. Recent experimental progress made it possible to locally
probe the formation of heavy quasi-particles in these systems via
scanning tunneling microscopy, in which strongly temperature-dependent
resonances at the Fermi energy have been observed. The shape of these
resonances varies
depending on the surface layer, i.e. if Cerium or Cobalt
terminates the sample. We clarify the microscopic origin of 
this difference
by taking into account the layered structure of the
material. Our simple model explains all the characteristic properties
observed experimentally, such as a layer-dependent shape of the
resonance at the Fermi energy, displaying a hybridization gap for
the Cerium-layer and a peak or dip structure for the other layers.
Our proposal resolves the seemingly unphysical assumptions in the 
preceding analysis based on the two-channel cotunneling model.
\end{abstract}


\maketitle
Recent advances in scanning tunneling microscopy (STM) provide the
ability to analyze strongly correlated materials with high spatial and
energy resolution, and thus enable to visualize
strongly correlated states. One of the more recent successes in the
field of STM has rendered it possible to examine heavy fermion materials.\cite{Aynajian2010,Schmidt2010,Hamidian2011,Ernst2010,Ernst2011,Aynajian2012,Maldonado2012,Allan2013,Zhou2013,roessler2013}
Before these STM measurements became possible, thermodynamic and transport measurements
showing e.g. a 
large effective electron mass in physical quantities, had been a main
probe for heavy fermion systems. 
This large effective electron mass can be 
theoretically explained by the formation of heavy quasi-particles, 
but a local probe to directly image these heavy quasi-particles was missing.

\textcite{Aynajian2012} studied CeMIn$_5$ (M=Co,
Rh) and succeeded in using the STM 
for locally observing the gradual formation of heavy quasi-particles, which
are ubiquitous for heavy fermion materials,
when lowering the temperature.
While the {\it f}-electrons in CeCoIn$_5$ strongly influence the material
properties, which give rise to an effective electron mass $10-50$ times of
the bare electron mass, the {\it f}-electrons in
CeRhIn$_5$ seem to be decoupled from the conduction
electrons. 
Thus, the comparison of both materials enables the authors to
distinguish the
influence of the strong correlations in the {\it f}-electrons.
Furthermore, the cleaving process used in these experiments leads to
the exposure of 
multiple termination surfaces which are distinguished by different chemical
components of the material. The STM spectra measured on the Ce-layer of
CeCoIn$_5$ reveal that around
the chemical potential a dip is formed, which is the signature
of the hybridization gap between {\it f}- and conduction-electrons. On
the other hand, spectra taken on the 
Co-surface of CeCoIn$_5$ show an enhancement of the spectral weight
for energies below the chemical potential
at low enough temperatures. 

These experimental results are explained by the authors
using the two-channel cotunneling model,\cite{Yang2009,Maltseva2009,Figgins2010,Wolfle2010} where it is assumed that the electrons of the
STM tip can simultaneously tunnel into the {\it f}-electron levels
(conduction electron levels) with the amplitude $t_f$ ($t_c$), respectively.  
By tuning the ratio $t_f/t_c$, they construct
 the spectra having a dip or a peak at the chemical potential. 
In order to explain the experimental findings, the STM spectra on
the Ce-layer (Co-layer)  are modeled by a large tunneling into the conduction
electron levels ({\it f}-electron levels).
However, as already noticed by \textcite{Aynajian2012}, this assumption
 seems  to be unphysical and is contrary to the general expectation that 
a tunneling into the strongly localized {\it f}-orbitals should be weak. 
Particularly, 
the claim of a strong tunneling from the STM tip into the {\it
  f}-electrons of Ce, when the STM tip is not located above Ce but
above Co, seems to be questionable.

In this paper, we propose an alternative and consistent explanation for the STM
spectra, which is based on the layered structure of CeCoIn$_5$, and 
 resolve the apparent inconsistency encountered in the previous 
analysis based on the two-channel cotunneling model. We will
qualitatively explain all the essential features in the experimental 
findings of \textcite{Aynajian2012} by assuming a tunneling only
between the STM-tip and the {\it spd}-conduction electrons, where
tunneling processes into the {\it f}-electron orbitals play an 
insignificant role on both of the Ce- and  Co-layers. Our scenario
demonstrates that the characteristic dip-peak structure in the STM profile
reflects the intrinsic properties of the material, which is contrasted to 
the previous scenario \cite{Aynajian2012}  where the interplay of the STM tip and the substrate would give such a structure.

{\it General idea :}  
The essential point in our scenario is
that CeCoIn$_5$ is built up of layers of different constituents;  
{\it f}-electrons only exist in the Ce-layers, while in the Co- and
In- layers only conduction electrons in {\it spd}-bands are
present. A key observation is that the Kondo effect in 
the {\it f}-orbitals of Ce, which leads
to the formation of heavy quasi-particles at low temperatures,
is not only observable in the orbitals of Ce, but also significantly 
influences the electrons of Co layers via the proximity effect 
between the Ce- and Co-layers. The penetration of the Kondo effect 
indeed results
in the formation of a strongly correlated resonance at 
the Fermi energy in the local density of states (LDOS) of 
Co-electrons. An important point is 
that the resulting resonance in the LDOS of Co-electrons
takes a different shape from that in the Ce-layer; due to the propagation
of electrons from the Ce-layer to the Co-layer this shape should be changed.

The mechanism proposed here for explaining the STM profile 
is based on the fact that CeCoIn$_5$ is a naturally-layered 
heavy fermion compound, implying that the observed dip-peak profile 
reflects the intrinsic properties of the layered material. In this
respect, the present scenario is quite different from that of the
two-channel cotunneling model \cite{Aynajian2012}, which relies on the 
interference effect between two possible tunneling processes to
produce such a structure.
We note here that in a recent related study we have analyzed
the paramagnetic state of artificially layered {\it f}-electron
superlattices, \cite{peters2013b} where we have shown that 
 the Kondo effect can be observed even in the conduction 
electrons of the spacer-layers in between the {\it f}-electron layers. 
Also,  in a slightly different context, such a layer-dependent shape 
of the Kondo resonance has been observed in STM spectra, 
which are taken 
on a Copper surface above a single magnetic impurity which is buried
below the surface.\cite{prueser2011} 
These recent studies naturally motivate us to 
demonstrate here that the STM measurements can probe the 
penetration of the Kondo effect of the Ce-{\it 
  f}-orbitals into the normal-metal layers at the surface of CeCoIn$_5$.

Another point we wish to stress here is that 
to explain the experimental findings
we do not need a fine tuning of the tunneling into {\it c}- and {\it
  f}-electron orbitals, $t_f$ and $t_c$, 
which has been previously used to reproduce the observed resonance.
We thus avoid an unrealistic strong tunneling \cite{Aynajian2012} 
between the STM-tip and the {\it f}-electron orbitals, 
which is expected to be weak, because {\it f}-electron orbitals are
strongly localized.

{\it Model and method : }
\begin{figure}[t]
\includegraphics[width=0.4\linewidth]{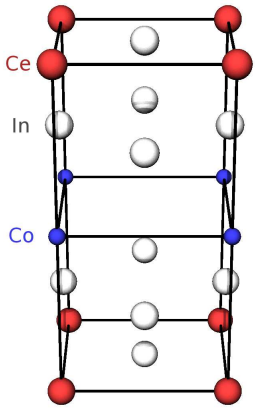}
\caption{(Color online) Crystal structure of CeCoIn$_5$.
 \label{cecoin}}
\end{figure}
We now demonstrate the above scenario by explicitly
performing the numerical calculation for a simplified model.
The crystal structure of CeCoIn$_5$, which is shown in
Fig. \ref{cecoin}, is built up of three different two-dimensional (2D)
layers. The Ce- and Co-atoms form a regular three-dimensional cubic
lattice, while the In-atoms occupy positions within the Ce-layers and
in layers which are located between the Ce- and Co-layers.
The measurements in ref.\cite{Aynajian2012} were taken on the Ce- and
the Co-layer. In order to simulate the STM spectra, we construct a
minimal model which is able to qualitatively reproduce the results.
We assume a single {\it spd}-orbital on each of the Ce-atoms
and the Co-atoms and a direct hopping between these two orbitals.
In order to include the effects induced by the strong
correlations in the {\it f}-orbitals of Ce, we furthermore couple
the conduction electrons of the Ce-atom to localized spins. 
We note that a more elaborate model for CeCoIn$_5$ should incorporate
  the multi-orbital structure of the {\it f}-electrons of Ce.
Our model thus consists of 
a regular cubic lattice, which is made of two different 2D layers: a square
lattice of Ce-atoms including {\it f}- and conduction-electrons 
({\it f}-electrons are approximated as localized spins), and the other
layers, which include only non-interacting electrons, corresponding to
the Co-layers.  
Furthermore, using open boundary conditions and
varying the number of layers in our system, we
are able to change  
 the nature of the surface layer in order to simulate the STM
spectra which are taken on various terminating layers of CeCoIn$_5$.
Thus, our theoretical model reads
\begin{eqnarray}
H&=&H_{Ce}+H_{0}+H_{inter}\nonumber\\
H_{Ce}&=&t\sum_{<i,j>z\sigma}c_{iz\sigma}^\dagger
c_{jz\sigma}+J\sum_{iz}\vec{s}_{iz}\cdot\vec{S}_{iz}\nonumber\\
H_{0}&=&t\sum_{<i,j>z\sigma}c_{iz\sigma}^\dagger
c_{jz\sigma}\nonumber\\
H_{inter}&=&t\sum_{i<z,z^\prime>\sigma}c_{iz\sigma}^\dagger
c_{iz^\prime\sigma}.
\end{eqnarray}
We assume that all hopping constants are equal for simplicity.
The correlation effects of the {\it f}-electrons are taken into
account by an antiferromagnetic coupling of the conduction electrons
to the localized spins in the Ce-layer,
$J\vec{s}_{iz}\cdot\vec{S}_{iz}=c_{iz\rho}^\dagger
c_{jz\rho^\prime}\vec{\sigma}_{\rho,\rho^\prime}\cdot \vec{S}_{iz}$
($\vec{\sigma}_{\rho,\rho^\prime}$: Pauli matrices).

We furthermore assume that the STM-tip can be modeled as a point
probe. The conductance between the {\it f}-electron material and the STM
tip is then proportional to the LDOS of the surface layer at the
location of measurement\cite{tersoff1985}. We will thus study in this
paper the surface LDOS of CeCoIn$_5$, which can be directly compared
to the STM spectra in reference.\cite{Aynajian2012}

For solving this model, we use the inhomogeneous dynamical mean
field theory (IDMFT)\cite{georges1996} in combination with the
numerical renormalization group (NRG).\cite{wilson1975,bulla2008}
Within DMFT, one describes the lattice model as a self-consistent
solution of a quantum impurity model. This approximation becomes exact
in the limit of infinite dimensions. However, in many previous
calculations it has been shown that DMFT 
can capture the essential physics of heavy fermion systems,
i.e. the low energy physics including heavy quasi-particles as well as
magnetic phases can be described by the
DMFT.\cite{jarrell1993,sun1993,rozenberg1995,sun2005,peters2007,leo2008,otsuki2009,hoshino2010,bodensiek2011,peters2012,peters2013} 
A detailed description of the IDMFT for superlattices can be found in reference.\cite{peters2013b}

\begin{figure}[t]
\includegraphics[width=1\linewidth]{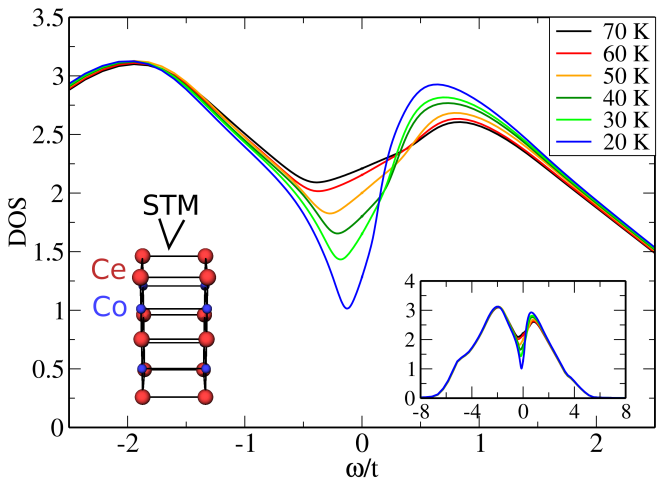}
\caption{(Color online) Temperature dependent LDOS close to the 
Fermi energy, $\omega=0$, for a system with a
  Ce-surface-layer. The temperatures are scaled so that the
  Kondo temperature of the simulated system is $50K$ corresponding to
  CeCoIn$_5$. The right inset shows the shape of the whole simulated
  band. The left inset shows the simplified lattice, we have used to
  simulate the spectra.\label{Ce-layer}}
\end{figure}
\begin{figure}[t]
\includegraphics[width=1\linewidth]{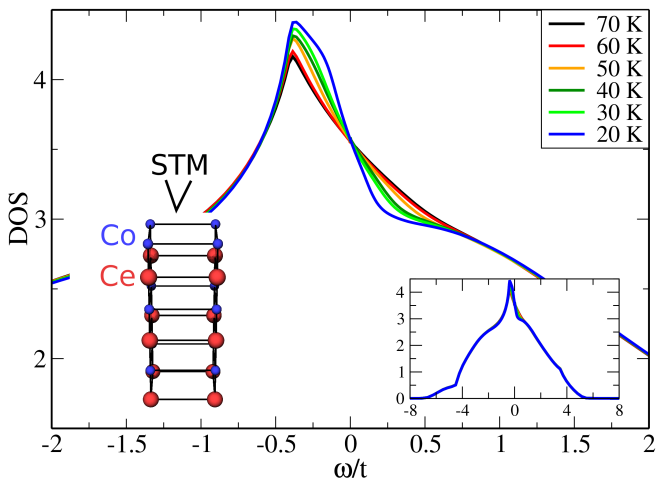}
\caption{(Color online) Same as Fig. \ref{Ce-layer}, but for a lattice
  corresponding to the Co-layer in CeCoIn$_5$. 
 \label{Co-layer}}
\end{figure}

{\it Simulated LDOS : } We simulate a slightly doped system with
$\langle n_{\text{Ce}}\rangle=1.1$ and $\langle
n_{\text{Co}}\rangle=1.2$ conduction electrons per Ce- and Co-atom,
respectively. We choose this doping in order to imitate
  the slope of the high 
temperature spectra of the experimental data, where the Kondo effect
does not influence the system. We believe that the asymmetries at
high temperatures in the experimental data are due to the lattice structure of
CeCoIn$_5$ and are not caused by the interplay of the
STM tip and the substrate of CeCoIn$_5$. By taking 
into account all Ce-, Co-, and In- energy bands which are close to the
Fermi energy, a better agreement between the simulated and
measured spectra at high temperatures can be expected. However, the
inclusion of all energy bands is out of reach in this
study. Nevertheless, our simplified model can qualitatively explain the
measured spectra.

We set the antiferromagnetic coupling between the
conduction electrons 
and the localized moments in the Ce-layers to $J=2t$, and treat systems
including up to $30$ layers. The shape of the Kondo
  resonance at the 
Fermi energy as well as the temperature dependence, 
when scaled with the Kondo temperature, do not depend on the coupling
strength.  The coherence temperature
for the formation of heavy quasi-particles for this interaction
strength is approximately $T_K\approx\frac{1}{5}t$. In order to compare
the temperature dependence of our results to the experimentally
measured STM spectra, we adjust our energy scale so that $T_K=50K$
roughly corresponding to the coherence temperature of CeCoIn$_5$.

In Figs. \ref{Ce-layer} and \ref{Co-layer}, we show the simulated LDOS of
the conduction electrons for the Ce- and Co-surface
layers. Fig. \ref{Ce-layer} corresponds to the Ce-layer including 
{\it f}-electrons. For the Ce-layer, a hybridization
gap opens at the Fermi energy when the temperature is lowered. The formation of
heavy quasi-particles by the hybridization between the {\it f}- and {\it
  spd}-electrons leads to a transfer of spectral weight from the Fermi
energy to energies above the Fermi energy in the LDOS of the
conduction electrons. The resulting pseudo-gap at the Fermi
energy as well as the peak above the Fermi energy and the temperature
dependence of these resonances agree
fairly well with the
measured STM spectra on the surface layer A in \textcite{Aynajian2012},
which has also been identified as Ce-layer. 

In Fig. \ref{Co-layer}, we show the LDOS of the non-interacting layer
which corresponds to the Co-layer of  CeCoIn$_5$ in our calculations.
A resonance due to the Kondo effect of the {\it f}-electrons in the
Ce-layer is also visible in this non-interacting layer. However, the
shape of this resonance is completely different.
Instead of a pseudo-gap at the Fermi energy, we
observe an enhanced spectral weight below the Fermi energy in the LDOS of the
conduction electrons. 
Furthermore, we observe that the
spectral weight above the Fermi energy is decreased when lowering the
temperature. 
The temperature
dependence of this spectral weight transfer is similar to that of the
formation of the hybridization gap in 
the Ce-layer. Around the coherence temperature, the spectral weight below
the Fermi energy suddenly increases. 
We want to stress that this enhanced spectral weight (peak) is observed in
the LDOS of the conduction electrons. We do not 
assume any direct tunneling into the {\it f}-electron levels. The
Kondo effect in the Ce-layers influences the LDOS of the
conduction electrons of the nearest 
neighboring layer. This resonance qualitatively agrees with
the STM spectra 
taken on the Co-surface of CeCoIn$_5$, where a peak is found slightly
below the Fermi energy.

We note that the results shown here are calculated via a
simplified crystal 
structure, i.e. a 3D cubic lattice instead of a realistic crystal
structure of CeCoIn$_5$ and neglecting the multi-orbital structure of
the {\it f}-electrons of Ce. The inclusion of the multi-orbital
structure will lead to further anisotropies in the shape of the
hybridization gap and a possible fine structure at finite
temperatures. Nevertheless, this simplified model is 
sufficient to qualitatively reproduce the experimentally observed 
resonances. 
 The main point is that the
resonances resulting from the Kondo effect of the Ce-atoms 
can be observed in the LDOS of the conduction electrons of the neighboring
non-interacting layers. The shape of these resonances at the Fermi energy
thereby depends on the distance between the probed layer and the {\it
  f}-electron layer.
If one uses a
realistic crystal structure, one can expect that the STM can also
quantitatively be reproduced by such calculations.

Before concluding the paper, some comments are in order on 
the theoretical background. The appearance of a Kondo resonance 
in the LDOS of the conduction electrons of
the non-interacting layers can be understood using the Dyson
equation. The LDOS, which is probed in STM measurements, corresponds to the
imaginary part of the interacting Green's function. The change in the
LDOS at
position $x$ and frequency $\omega$  originating from correlation
effects in the Ce-layer, can be 
written as
\begin{equation}
\Delta\rho(x,\omega)=-\frac{1}{\pi}Im\left(G^0(x,x_{Ce},\omega)T(x_{Ce},\omega)G^0(x_{Ce},x,\omega) \right)\nonumber,
\end{equation}
where $G^0(x,x_{Ce})$ is the free Green's function for the
non-interacting conduction electrons. All interaction effects
of the {\it f}-electrons are included in the $T$-matrix,
$T(x_{Ce},\omega)$, which for the case of heavy fermions develops a
resonance at the Fermi energy below the coherence temperature. This
resonance is carried via the non-interacting electrons from the
Ce-atoms, $x_{Ce}$, to the place of the STM measurements, $x$.

{\it Conclusions :}
We have analyzed the surface LDOS of layered heavy fermion systems such as
CeCoIn$_5$. We have shown that resonances originating from the Kondo
effect of the {\it f}-electrons occur at the Fermi energy. These
resonances can be observed not only in the conduction electrons of the
Ce-atoms, but also in the LDOS of the conduction electrons of all
other layers, such as the Co- and the
In-layers. The shape of this resonance is quite different for different
layers and depends on the distance from the {\it f}-electron
layer. Therefore, the STM results in \textcite{Aynajian2012} can be
unambiguously 
explained by tunneling into the conduction electrons of different
layers which have different distance to the Ce-layer. For explaining
the measured spectra, we do not need a fine tuning done
 by \textcite{Aynajian2012} 
 for separate tunneling into the {\it c}- and {\it   f}-electrons. 
Because {\it f}-orbitals are strongly
localized, a direct tunneling into these orbitals should be very
small and would be negligible in the Co-layers.
Taking a realistic crystal structure of CeCoIn$_5$ or other heavy
fermion materials into account, we
believe that measured STM spectra can be quantitatively
reproduced. This is left as a future project.

\begin{acknowledgments}
We thank H. Ikeda for useful discussions.
RP and NK thank the Japan Society for the Promotion of Science (JSPS)
for the support  through its FIRST Program. 
NK acknowledges support through KAKENHI (No. 22103005, No. 25400366).
The numerical 
calculations were performed at the ISSP 
in Tokyo and on the SR16000 at  YITP in Kyoto University.
\end{acknowledgments}

\end{document}